\magnification=1200
\def\qed{\unskip\kern 6pt\penalty 500\raise -2pt\hbox
{\vrule\vbox to 10pt{\hrule width 4pt\vfill\hrule}\vrule}}
\centerline
{A MECHANICAL MODEL FOR FOURIER'S LAW OF HEAT CONDUCTION.}
\bigskip\bigskip\bigskip\bigskip
\centerline{by David Ruelle\footnote{$\dagger$}{Math. Dept., Rutgers University, and 
IHES, 91440 Bures sur Yvette, France. email: ruelle@ihes.fr}.}
\bigskip\bigskip\bigskip\bigskip\noindent
	{\leftskip=2cm\rightskip=2cm\sl Abstract. Nonequilibrium statistical mechanics close to equilibrium is a physically satisfactory theory centered on the linear response formula of Green-Kubo.  This formula results from a formal first order perturbation calculation without rigorous justification.  A rigorous derivation of Fourier's law for heat conduction from the laws of mechanics remains thus a major unsolved problem.  In this note we present a deterministic mechanical model of a heat-conducting chain with nontrivial interactions, where kinetic energy fluctuations at the nodes of the chain are removed.  In this model the derivation of Fourier's law can proceed rigorously.\par}
\vfill\eject
\noindent
{\bf 0. Introduction.}
\medskip
	To understand the {\it transport phenomena} of nonequilibrium thermodynamics from the point of view of microscopic dynamics (say classical mechanics) is a serious challenge.  Formally, this is a problem of {\it linear response}, solved by the Green-Kubo formula, which is basically the result of a first order perturbation calculation.  But this perturbation calculation is an uncontrolled approximation, as remarked by van Kampen [22].  A fundamental derivation of Fourier's law for heat conduction remains thus an open problem, as repeatedly pointed out by Lebowitz (see for instance Bonetto et al. [4]).
\medskip
	Let $\rho_0$ be a microcanonical equilibrium state, which is an invariant probability measure for the microscopic dynamics of the physical system of interest.  The linear response problem involves finding the {\it physical state} $\rho$ which replaces $\rho_0$ when a small change is made to the microscopic dynamics.  Problems of this type are mathematically well understood when the ``microscopic dynamics'' corresponds to {\it uniformly hyperbolic} smooth dynamics on a compact manifold $M$ and $\rho$ is a so-called SRB state on $M$\footnote{(*)}{There is a large literature on this subject.  See for instance Young [23] for a general discussion of SRB (Sinai, Ruelle, Bowen) states.  Linear response for uniformly hyperbolic diffeomorphisms was established by Katok et al. [16], see also Ruelle [18].  The flow case, which is most relevant for physics, is discussed in Ruelle [22], Butterley and Liverani [5] (For applications to physics, see Gallavotti and Cohen [13], Ruelle [19]).  Further discussion and references can be found in Dolgopyat [8] who discusses extensions to partially hyperbolic dynamics.  An idea of the variety of nonhyperbolic dynamics is given by Bonatti et al. [3].  See also in this respect Baladi and Smania [2], and Ruelle [21].}.  Uniform hyperbolicity is however too strong a requirement in the physical situation of interest here.  In fact, as a consequence of the spatial extension of our physical system, there appear a number of ``central directions'' for the dynamics, i.e., we have to deal with {\it partially hyperbolic} dynamics.  We shall see how this happens in a model discussed below, and how some results of Dolgopyat [8] can be applied to a situation where the dynamics is partially hyperbolic, and SRB states are replaced by $u$-Gibbs states.
\medskip
	To be specific, the purpose of this note is to discuss a deterministic mechanical model which exhibits realistic behavior for heat conduction.  To obtain our model we start with a Hamiltonian chain of $N+1$ nontrivially coupled mechanical systems (nodes), and we fix the temperatures $T_0,T_N$ of the endpoints of the chain.  The kinetic energies of the intermediate nodes fluctuate.  In our model we remove the fluctuations by thermostats that fix the intermediate temperatures.  We define a {\it stable temperature profile} by requiring that the intermediate temperatures be such that, for each thermostat, there is no net flux of energy in or out of the corresponding node.  If this requirement is not satisfied we expect the intermediate temperatures to move towards the stable temperature profile when the thermostats are removed (see Section 4).  Our model allows thus to determine the intermediate temperatures for a heat-conducting chain.  Removing the kinetic energy fluctuations of the nodes of the original Hamiltonian chain is an uncontrolled approximation, but once it is accepted one can proceed rigorously.  One of our results will be {\it Fourier's Law}: the amount of energy transported by the chain is asymptotically $\sim N^{-1}(T_N-T_0)$ for large $N$ and small $T_N-T_0$ (see Section 5 for a more prudent statement).
\medskip
	The treatment given here strives at conceptual clarity rather than generality.  At the cost of increased mathematical complexity (or new ideas) one could probably deal with much more general situations than the simple model discussed below.  Note that there are a number of rigorous papers related to the statistical mechanics of heat conduction, but using approaches different from that presented here.  This includes work by Eckmann, Gallavotti, Hairer, Jak\v si\'c, Liverani, Pillet, Rey-Bellet, Young, etc. (see in particular [9], [1], [10], [11]).  See also a study with stochastic thermostats [4a], [4b], and a promising investigation by Dolgopyat and Liverani [8a] of the macroscopic behavior of a coupled lattice of strongly chaotic microscopic subsystems.
\medskip\noindent
{\bf Acknowledgments.}
\medskip
	I am very thankful to Dmitry Dolgopyat for clarifying for me a critical point of his paper [8].  I am also indebted for a number of discussions on linear response in smooth dynamics and nonequilibrium statistical mechanics to Viviane Baladi, Jean-Pierre Eckmann, Giovanni Gallavotti, Vojkan Jak\v si\'c,  Joel Lebowitz, and Lai-Sang Young.
\medskip\noindent
{\bf 1. Our model: the time evolution $(f^t)$.}
\medskip
	Our model is a chain of $N+1$ nodes.  The nodes, before coupling, are assumed to be Hamiltonian systems described by geodesic flows on a compact $n$-dimensional Riemann manifold $M$, with $n\ge2$.  The Hamiltonian $H_j$ of the $j$-th node is thus the kinetic energy.  In local coordinates
$$	H_j={1\over2m}\langle{\bf p}_j,{\bf p}_j\rangle   $$
where we have written ${\bf p}_j=(p_{ju})$, ${\bf q}_j=(q_j^u)$, $\langle a,b\rangle=\sum_{uv}g^{uv}({\bf q}_j)a_ub_v$, and $(g^{uv})$ is the inverse of the matrix $(g_{uv})$ defining the metric (the nodes have mass $m$).  This choice of Hamiltonian system gives examples where the time evolution is an exponentially mixing Anosov flow.  Furthermore there will be a simple relation between the kinetic energy and the temperature (see Section 4).
\medskip
	We write ${\bf x}_j=({\bf p}_j,{\bf q}_j)$ and let $(f_j^t)$ be the geodesic flow $({\bf x}_j,t)\mapsto f_j^t{\bf x}_j$ restricted to the {\it energy shell} $S_j=\{{\bf x}_j:H_j({\bf x}_j)=K_j\}$ with $K_j>0$.  Defining $S=\times_{j=0}^NS_j\subset T^*(M^{N+1})$, we also let $f_\times^t=\times_{j=0}^Nf_j^t$ on $S$.  The time evolution defined by the Hamiltonian $\sum_0^NH_j$, when restricted to $S$, is thus $(f_\times^t)$. 
\medskip
	We introduce now a coupling between the node $j$ and its neighbors, given by a force $\lambda X_j\in T_{{\bf q}_j}^*M$ where $\lambda\in{\bf R}$.  We require that there is a smooth function $W:M\times M\to{\bf R}$ [satisfying condition (1.3) below] such that
$$	X_j=X_j^-+X_j^+   $$
where $X_0^-=X_N^+ =0$, and the other $X_j^\pm$ are given by
$$	X_j^-=-\partial_{{\bf q}_j}W({\bf q}_{j-1},{\bf q}_j)\qquad,\qquad
	X_j^+=-\partial_{{\bf q}_j}W({\bf q}_j,{\bf q}_{j+1})\eqno{(1.1)}   $$
[For simplicity we do not introduce a self-force $X_j^0$ depending only on ${\bf q}_j$.]  
The time evolution corresponding to the coupled Hamiltonian
$$     H=\sum_{j=0}^N{1\over2m}\langle{\bf p}_j,{\bf p}_j\rangle
	+\lambda\sum_{j=1}^NW({\bf q}_{j-1},{\bf q}_j)   $$
is given by
$$	{d\over dt}\pmatrix{p_{ju}\cr q_j^u\cr}
=\pmatrix{-\sum_{vw}(\partial g^{vw}/\partial q_j^u)p_{jv}p_{jw}/2m+\lambda X_{ju}\cr
	\sum_v g^{uv}({\bf q}_j)p_{jv}/m\cr}   $$
\medskip
   Write now
$$	\hat\alpha_j=\langle X_j,{\bf p}_j\rangle
	=\hat\alpha_j^-+\hat\alpha_j^+\quad{\rm where}\quad
	\hat\alpha^\pm=\langle X_j^\pm,{\bf p}_j\rangle   $$
$$	\alpha_j=\hat\alpha_j/\langle{\bf p}_j,{\bf p}_j\rangle
	=\alpha_j^-+\alpha_j^+\quad{\rm where}\quad
	\alpha_j^\pm=\hat\alpha_j^\pm/\langle{\bf p}_j,{\bf p}_j\rangle   $$ 
We specify our model to correspond to the following coupled time evolution $(f^t)$ on $S$ for the $N+1$ nodes:
$$	{d\over dt}\pmatrix{p_{ju}\cr q_j^u\cr}
	=\pmatrix{-\sum_{vw}(\partial g^{vw}/\partial q_j^u)p_{jv}p_{jw}/2m
	+\lambda X_{ju}-\lambda\alpha_jp_{ju}\cr
	\sum_v g^{uv}({\bf q}_j)p_{jv}/m\cr}\eqno{(1.2)}   $$
The choice of the $\alpha_j$ is such that for the coupled time evolution we have
$$	{d\over dt}H_j={1\over m}\sum_{uv}g^{uv}({\bf q}_j)p_{ju}{d\over dt}p_{jv}
+{1\over2m}\sum_{uv}\partial g^{uv}({\bf q}_j)/\partial q_j^wp_{ju}p_{jv}{d\over dt}q_j^w
	=0   $$
for $j=0,\ldots,N$.  The term $-\lambda\alpha_jp_{ju}$ in $(1.2)$ is called an {\it isokinetic thermostat} (introduced by Evans and Hoover, see [12], [15]): it keeps the kinetic energy of node $j$ fixed to a value $K_j$ for each $j$.  [This is a physically reasonable thermostat, especially when $n$ is large.]  Therefore $(f^t)$ is a time evolution on $S$ as announced.  (We shall complete the description of our model by making a specific choice of the kinetic energies $K_j$, see the definition of a stable temperature profile in Section 5.)
\medskip
	Let us define
$$	\rho_j(d{\bf x}_j)=d{\bf p}_j\,d{\bf q}_j   $$
where $d{\bf p}_j$ is the normalized volume on the sphere $\{{\bf p}_j:\sum_{uv}g^{uv}({\bf q}_j)p_{ju}p_{jv}=2mK_j\}$, and $d{\bf q}_j$ is the normalized Riemann volume on $M$; $\rho_j$ is thus an an ergodic measure for $(f_j^t)$ on $S_j$.  We also write ${\bf x}=({\bf x}_0,\ldots,{\bf x}_N)$, and define $\rho_\times(d{\bf x})=\prod_{j=0}^N\rho_j(d{\bf x}_j)$.  We assume that
$$	\int d{\bf q}_{j-1}W({\bf q}_{j-1},{\bf q}_j)
	=\int d{\bf q}_jW({\bf q}_{j-1},{\bf q}_j)=0\eqno{(1.3)}   $$
Note that, since $d{\bf p}_j$ is invariant under ${\bf p}_j\mapsto-{\bf p}_j$, we have
$$	\rho_\times(\hat\alpha_j^\pm)=0\eqno{(1.4)}   $$
Furthermore,
$$	\rho_\times((\hat\alpha_j^+)(\hat\alpha_k^\epsilon\circ f_\times^t))=0\quad\hbox
	{unless}\quad\hbox{$\hat\alpha_k^\epsilon$
	is $\hat\alpha_j^+$ or $\hat\alpha_{j+1}^-$}\eqno{(1.5)}    $$
$$	\rho_\times((\hat\alpha_j^-)(\hat\alpha_k^\epsilon\circ f_\times^t))=0\quad\hbox
	{unless}\quad\hbox{$\hat\alpha_k^\epsilon$
	is $\hat\alpha_j^-$ or $\hat\alpha_{j-1}^+$}\eqno{(1.6)}    $$
To see this, note that $f_\times^t$ does not mix different ${\bf x}_\ell$, that $\rho_\times$ is $(f_\times^t)^*$-invariant, that $\int d{\bf p}_\ell\langle X_\ell^\pm,{\bf p}_\ell\rangle=0$ (invariance of $d{\bf p}_\ell$ under ${\bf p}_\ell\mapsto-{\bf p}_\ell$), and that $\int d{\bf q}_{\ell\pm1}\langle X_\ell^\pm,{\bf p}_\ell\rangle=0$ (because of (1.1) and (1.3)).
\medskip\noindent
{\bf 2. A perturbation argument.}
\medskip
	Since ${\bf \rho}_j$ has a smooth density on $S_j$, we see that it is an SRB state for $(f_j^t)$, and also that $\rho_\times$ is an SRB state for $(f_\times^t)$ on $S$.  {\sl We shall now assume that the $(f_j^t)$ are exponentially mixing Anosov flows}.  (Since $(f_j^t)$ is the geodesic flow on a compact Riemann manifold $M$, this is the case if $M$ is a surface of negative curvature, see [6], [7].)  We refer the reader to Appendix A for a precise definition and a proof that $(\rho_\times,(f_\times^t))$ is also exponentially mixing.  In particular $(\rho_\times,(f_\times^t))$ is ergodic.
\medskip
   To study the physics corresponding to the perturbed time evolution $(f^t)$ defined by $(1.2)$, the existence of an SRB state $\rho$ would be desirable.  But, since $(f^t)$ is not uniformly hyperbolic, such a state need not exist when $\lambda\ne0$.  It is however possible, following an idea of Dolgopyat [8], to study a perturbation of $\rho_\times$ to $u$-Gibbs states, as we now explain.
\medskip
   A dynamical system $(f^t)$ on the compact Riemann manifold $S$ is said to be {\it partially hyperbolic} if there is a continuous invariant splitting
$$   TS=E^u\oplus E^{cs}   $$
such that, for suitable constants $C>0$, $\theta_1,\theta_2<1$, and all $t\ge0$, we have
$$   (\forall v\in E^u)\qquad ||(Tf^{-t})v||\le C\theta_1^t||v||   $$
$$   (\forall x\in S)\qquad ||(Tf^t)|E_x^{cs}||/||(Tf^{-t})|E_x^u||^{-1}\le C\theta_2^t   $$
One can then define local {\it unstable} manifolds ${\cal W}_x^u$; the corresponding global manifolds form a continuous foliation ${\cal W}^u$ of $S$ with smooth leaves, which is tangent to $E^u$ (see [14]).
\medskip
   An $(f^t)$-invariant probability measure $\rho$ on $S$ is called (by Pesin and Sinai [17]) a $u$-{\it Gibbs state} if the conditional measures on the local unstable manifolds have a density of a certain canonical form.  The $u$-Gibbs states are precisely the $(f^t)$-invariant probability measures $\rho$ on $S$ which are absolutely continuous with respect to the foliation ${\cal W}^u$ (i.e., if $X\cap{\cal W}_x$ has leaf Lebesgue measure 0 for each local unstable manifold ${\cal W}_x$, then $\rho(X)=0$).  See Dolgopyat [8] for a discussion of $u$-Gibbs states from this point of view.  If $\rho$ is SRB, then $\rho$ is also $u$-Gibbs.  If $\ell$ is a probability measure absolutely continuous with respect to the Riemann volume on $S$, and
$$   \ell_T={1\over T}\int_0^Tdt(f^t)^*\ell   $$
then any weak limit of $\ell_T$ when $T\to\infty$ is a $u$-Gibbs state.  In particular, the set of $u$-Gibbs states is nonempty.
\medskip
	A $u$-Gibbs state, while $(f^t)$-invariant, may have a natural decomposition into states corresponding to slow time-oscillations of our physical system.  This possibility makes $u$-Gibbs states more flexible objects than SRB states for the description of nonequilibrium steady states.
\medskip
   In the physical situation that we want to discuss, $(f_\times^t)$ is partially hyperbolic because the $(f_j^t)$ are Anosov flows.  The perturbed time evolution $(f^t)$ is thus also partially hyperbolic, provided $\lambda$ is small enough (see [17]).  For $\lambda\ne0$ there need not be an SRB state, but limits of $\ell_T$ are $u$-Gibbs states, and the following applies:
\medskip\noindent
{\bf 3. Proposition.}
\medskip
   {\sl For $\lambda$ sufficiently small, if $\rho$ is a $u$-Gibbs state for $(f^t)$, and $A$ a smooth function on $S$, we have
$$   \rho(A)-\rho_\times(A)=\lambda(n-1)\int_0^\infty d\tau\rho_\times((\sum_{j=0}^N\alpha_j)(A\circ f_\times^\tau))+o(\lambda)\eqno{(3.1)}   $$}
\medskip
   This result is a corollary of a theorem of Dolgopyat [8], which applies to the time 1 map $f_\times^1$ because of our assumptions.  Specifically, $f_\times^1$ is a rapidly mixing (in fact exponentially mixing) Anosov element in an abelian Anosov action on S.  Therefore, for sufficiently small $\lambda$, if $\rho$ is a $u$-Gibbs state for $(f^t)$ (hence $f^1$), Theorem 1 of [8] yields
$$   \rho(A)-\rho_\times(A)=\lambda\omega(A)+o(\lambda)\eqno{(3.2)}   $$
where $A\mapsto\omega(A)$ is a linear functional on smooth functions.  We refer to [8] for the definition of abelian Anosov action and other details including weaker conditions on $A$.  A proof that $f_\times^1$ is rapidly mixing in the sense of [8] is given in Appendix A.
\medskip
	Dolgopyat in [8] gives an explicit expression for the linear functional $\omega$.  Since $f_\times^1$ preserves the volume element $\rho_\times(d{\bf x})$ on $S$, Dolgopyat's expression can be simplified to
$$	\omega(A)
	=-\sum_{k=0}^\infty\rho_\times([{\rm div}Y][A\circ f_\times^k])\eqno{(3.3)}$$
where the divergence is taken with respect to $\rho_\times$, and
$$	Y=\big({df^1\over d\lambda}\circ f^{-1}\big)\big|_{\lambda=0}   $$
We refer to Appendix B for the proof of (3.3).  Finally, we can compute ${\rm div}Y$ in our case; this is done in Appendix C and yields (3.1).\qed
\medskip\noindent
{\bf 4. Energy transfers.}
\medskip
	Taking $A=\hat\alpha_j^\pm$ in $(3.1)$, and using $(1.4-1.6)$, we have
$$	\rho(\hat\alpha_j^\pm)=\lambda(n-1)\int_0^\infty d\tau(\rho_j\rho_{j\pm1})
	[(\alpha_j^\pm+\alpha_{j\pm1}^\mp)(\hat\alpha_j^\pm\circ f_\times^\tau)]
	+o(\lambda)   $$
Since on $S_j$ we have $\hat\alpha_j^\pm=2mK_j\alpha_j^\pm$, we obtain
$$	\rho(\hat\alpha_j^\pm)={\lambda\over m}\int_0^\infty d\tau(\rho_j\rho_{j\pm1})
	[(\beta_j\hat\alpha_j^\pm+\beta_{j\pm1}\hat\alpha_{j\pm1}^\mp)
	(\hat\alpha_j^\pm\circ f_\times^\tau)]+o(\lambda)\eqno{(4.1)}   $$
where $\beta_j^{-1}=2K_j/(n-1)$ is defined to be the temperature associated with the kinetic energy $K_j$.
\medskip
	Using the invariance of $\rho_j$ under $({\bf p},{\bf q})\mapsto(-{\bf p},{\bf q})$ we have
$$	\int_0^\infty d\tau(\rho_j\rho_{j-1})
	[(\hat\alpha_{j-1}^+)(\hat\alpha_j^-\circ f_\times^\tau)]
	=\int_0^\infty d\tau(\rho_j\rho_{j-1})
	[(\hat\alpha_{j-1}^+)(\hat\alpha_j^-\circ f_\times^{-\tau})]   $$
$$	=\int_{-\infty}^0 d\tau(\rho_j\rho_{j-1})
	[(\hat\alpha_{j-1}^+)(\hat\alpha_j^-\circ f_\times^\tau)]   $$
and similarly with interchange of $+$ and $-$, so that $(4.1)$ becomes
$$	\rho(\hat\alpha_j^\pm)
	={\lambda\over2m}\int_{-\infty}^\infty d\tau(\rho_j\rho_{j\pm1})
	[(\beta_j\hat\alpha_j^\pm+\beta_{j\pm1}\hat\alpha_{j\pm1}^\mp)
	(\hat\alpha_j^\pm\circ f_\times^\tau)]+o(\lambda)\eqno{(4.2)}   $$
\par
	Let $(\phi^t)$ be the geodesic flow with unit velocity on $M$, and $\rho_\phi$ the corresponding absolutely continuous invariant measure.  Writing $W({\bf x}_1,{\bf x}_2)$ instead of $W({\bf q}_1,{\bf q}_2)$ we define
$$	\Psi(p_1,p_2,\tau_1-\sigma_1,\tau_2-\sigma_2)
	\qquad\qquad\qquad\qquad\qquad\qquad\qquad\qquad   $$
$$	\qquad\qquad=\int\rho_\phi(d{\bf x}_1)\rho_\phi(d{\bf x}_2)
	W(\phi^{p_1\sigma_1}{\bf x}_1,\phi^{p_2\sigma_2}{\bf x}_2)
	W(\phi^{p_1\tau_1}{\bf x}_1,\phi^{p_2\tau_2}{\bf x}_2)   $$
$$	\Psi_{ij}(p_1,p_2,\tau_1-\sigma_1,\tau_2-\sigma_2)
=\partial_{\sigma_i}\partial_{\tau_j}\Psi(p_1,p_2,\tau_1-\sigma_1,\tau_2-\sigma_2)   $$
and
$$	\Phi_{ij}(p_1,p_2)=\int_{-\infty}^{+\infty}d\tau\,\Psi_{ij}(p_1,p_2,\tau,\tau)   $$
Then we may write $\Phi=\Phi_{11}=-\Phi_{12}=-\Phi_{21}=\Phi_{22}\ge0$.  Because of the exponential decay of correlations, $\Phi(p_1,p_2)$ depends smoothly on $p_1,p_2$ $>0$.  Note also that
$$	\Phi(p_1,p_2)=\Phi(p_2,p_1)\qquad,\qquad\Phi(p,p)=p\Phi(1,1)   $$
It is easy to check that
$$	\int_{-\infty}^{+\infty}d\tau\,(\rho_{j-1}\rho_j)[(\hat\alpha_{j-1}^+)
	(\hat\alpha_{j-1}^+\circ f_\times^\tau)]=\Phi(|{\bf p}_{j-1}|,|{\bf p}_j|)   $$
$$	\int_{-\infty}^{+\infty}d\tau\,(\rho_{j-1}\rho_j)[(\hat\alpha_{j-1}^+)
	(\hat\alpha_j^-\circ f_\times^\tau)]=-\Phi(|{\bf p}_{j-1}|,|{\bf p}_j|)   $$
$$	\int_{-\infty}^{+\infty}d\tau\,(\rho_{j-1}\rho_j)[(\hat\alpha_j^-)
	(\hat\alpha_{j-1}^+\circ f_\times^\tau)]=-\Phi(|{\bf p}_{j-1}|,|{\bf p}_j|)   $$
$$	\int_{-\infty}^{+\infty}d\tau\,(\rho_{j-1}\rho_j)[(\hat\alpha_j^-)
	((\hat\alpha_j^-\circ f_\times^\tau)]=\Phi(|{\bf p}_{j-1}|,|{\bf p}_j|)   $$
In particular, $(4.2)$ gives
$$	\rho(\hat\alpha_j^-)
={\lambda\over2m}(\beta_j-\beta_{j-1})\Phi(|{\bf p}_{j-1}|,|{\bf p}_j|)+o(\lambda)   $$
\par
	By definition of $\hat\alpha_j^\pm$, the average total transfer of energy per unit time from the node $j-1$ to the node $j$ is thus
$$	{\lambda\over m}[\rho(\hat\alpha_j^-)-\rho(\hat\alpha_{j-1}^+)]
	={\lambda^2\over m^2}(\beta_j-\beta_{j-1})\Phi(|{\bf p}_{j-1}|,|{\bf p}_j|)
	+o(\lambda^2)\eqno{(4.3)}   $$
for $j=1,\ldots,N$.  Note in this respect that
$$	\rho(\hat\alpha_{j-1}^++\hat\alpha_j^-)=0   $$
This is because
$$	\hat\alpha_{j-1}^++\hat\alpha_j^-
	=\langle X_{j-1}^+,{\bf p}_{j-1}\rangle+\langle X_j^-,{\bf p}_j\rangle   $$
$$	=(\partial_{t_1}+\partial_{t_2})W({\bf q}_{j-1}(t_1),{\bf q}_j(t_2))\big|_{t_1=t_2=t}
	=\partial_tW({\bf q}_{j-1}(t),{\bf q}_j(t))   $$
and the average of the right-hand side over the $(f^t)$-invariant measure $\rho$ vanishes.
\medskip
	The average transfer of energy per unit time from the thermostatting force $-\lambda(\alpha_j^-+\alpha_j^+){\bf p}_j/m$ to the node $j$ can be obtained by a similar calculation.  It is
$$	-{\lambda\over m}\rho(\hat\alpha_j^-+\hat\alpha_j^+)   $$
$$	=-{1\over2}{\lambda^2\over m^2}[(\beta_j-\beta_{j-1})\Phi(|{\bf p}_j|,|{\bf p}_{j-1}|)
+(\beta_j-\beta_{j+1})\Phi(|{\bf p}_j|,|{\bf p}_{j+1}|)]+o(\lambda^2)\eqno{(4.4)}   $$
This is also minus the initial rate of heating of the node $j$ in the absence of a thermostat.
\medskip\noindent
{\bf 5. Stable and approximate stable temperature profiles.}
\medskip
	It is natural to define a {\it stable temperature profile} (STP) by fixing $\beta_0$ and $\beta_N$, and requiring $\beta_1,\ldots,\beta_{N-1}$ to be such that $\rho(\hat\alpha_j^-+\hat\alpha_j^+)=0$ for $j=1,\ldots,N-1$.  [Equivalently, $K_0$, $K_N$ are fixed, and $K_1,\ldots,K_{N-1}$ are such that $\rho(\hat\alpha_j^-+\hat\alpha_j^+)=0$.]  In an STP, there is thus no net energy contribution from the thermostats to the nodes $1,\ldots,N-1$, but the thermostats remove the kinetic energy fluctuations at these nodes.
\medskip
	We expect that, for given $\beta_0,\beta_N$, some STP exists, at least when $\lambda$ is sufficiently small, but we do not have a proof of that fact.  Technically, what is lacking for a rigorous discussion of STP's is a proof of uniformity of $o(\lambda)$, in $(3.1)$ or $(4.1)$, with respect to $\beta_0,\beta_1,\ldots,\beta_N$ in a compact interval\footnote{*}{This uniformity can probably be proved, according to Dmitry Dolgopyat (private communication).}
\medskip
	This being the case, we shall content ourselves with a discussion of {\it approximate} STP's defined by
$$	(\beta_j-\beta_{j-1})\Phi(|{\bf p}_{j-1}|,|{\bf p}_j|)
	=(\beta_{j+1}-\beta_j)\Phi(|{\bf p}_j|,|{\bf p}_{j+1}|)\eqno{(5.1)}   $$
for $j=1,\ldots,N-1$.  For an approximate STP, $(4.4)$ says that the average energy transfer per unit time from the thermostatting forces is $o(\lambda^2)$, which is very small for small $\lambda$.  The total energy transfer per unit time through our chain, i.e., $\rho(\hat\alpha_N^-)-\rho(\hat\alpha_0^+)$ is thus, according to $(4.3)$
$$	={\lambda^2\over m^2}(\beta_j-\beta_{j-1})\Phi(|{\bf p}_{j-1}|,|{\bf p}_j|)
	+o(\lambda^2)   $$
for $j=1,\ldots,N$.
\medskip
	We take now for definiteness $\beta_0<\beta_N$.  We know that $\Phi(p_1,p_2)\ge0$, and we shall assume that
$$	0<\Phi_{\rm min}\le\Phi(p_1,p_2)\le\Phi_{\rm max}   $$
when $p_1,p_2$ belong to some compact set where the $|{\bf p}_j|$ are allowed to vary: the condition $\Phi_{\rm min}>0$ expresses that the nodes of our chain are actually interacting.  We may then rewrite $(5.1)$ as
$$	\beta_j={\Phi^-\over\Phi^-+\Phi^+}\beta_{j-1}
	+{\Phi^+\over\Phi^-+\Phi^+}\beta_{j+1}   $$
where $\Phi^-=\Phi(|{\bf p}_{j-1}|,|{\bf p}_j|)$, $\Phi^+=\Phi(|{\bf p}_j|,|{\bf p}_{j+1}|)$.  We have thus $\beta_0<\beta_1<\ldots<\beta_N$, and also
$$	(\beta_j-\beta_{j-1})\Phi_{\rm min}
	\le(\beta_j-\beta_{j-1})\Phi(|{\bf p}_{j-1}|,|{\bf p}_j|)   $$
$$	={1\over N}\sum_{j=1}^N(\beta_j-\beta_{j-1})\Phi(|{\bf p}_{j-1}|,|{\bf p}_j|)
	\le{1\over N}(\beta_N-\beta_0)\Phi_{\rm max}   $$
Therefore
$$	0<\beta_j-\beta_{j-1}
	\le{1\over N}(\beta_N-\beta_0)\Phi_{\rm max}/\Phi_{\rm min}   $$
tends to zero when $N\to\infty$.
\medskip
	If $\kappa(N)=(\lambda^2/m^2)(\beta_j-\beta_{j-1})\Phi(|{\bf p}_{j-1}|,|{\bf p}_j|)$ is the total energy transfer per unit time through the chain (up to $o(\lambda^2)$) we have thus for large $N$
$$\kappa(N)\approx{\lambda^2\over m^2}(\beta_j-\beta_{j-1})\Phi(|{\bf p}_j|,|{\bf p}_j|)
	={\lambda^2\over m^2}(\beta_j-\beta_{j-1})|{\bf p}_j|\Phi(1,1)   $$
hence
$$	\Delta\beta_j=\beta_j-\beta_{j-1}
	\approx{m^2\kappa(N)\over\lambda^2\Phi(1,1)}{1\over|{\bf p}_j|}
	={m^2\kappa(N)\over\lambda^2\Phi(1,1)}{\beta_j^{1/2}\over\sqrt{m(n-1)}}   $$
$$	\Delta\beta_j^{1/2}={1\over2}{\Delta\beta_j\over\beta_j^{1/2}}
	\approx{1\over2}{m^2\kappa(N)\over\lambda^2\Phi(1,1)\sqrt{m(n-1)}}   $$
so that
$$	\beta_N^{1/2}-\beta_0^{1/2}
	\approx{N\over2}{m^2\kappa(N)\over\lambda^2\Phi(1,1)\sqrt{m(n-1)}}   $$
Finally
$$	\kappa(N)\approx{1\over N}
	\cdot{2\lambda^2\Phi(1,1)\over m^2}\sqrt{m(n-1)}(\beta_N^{1/2}-\beta_0^{1/2})   $$
conforms to Fourier's law:
$$	\kappa(N)\sim{\beta_0^{-1}-\beta_N^{-1}\over N}   $$
when $N$ is large and $\beta_N-\beta_0$ small.
\medskip\noindent
{\bf Appendix A: exponential mixing and rapid mixing.}
\medskip
	Let $(f_j^t)$ be a smooth flow on the compact manifold $S_j$.  We say that $(f_j^t)$ is {\it exponentially mixing} with respect to the invariant state $\rho_j$ if for some $p>0$ there are $\gamma,C>0$, such that
$$	|\rho_j((A_j\circ f_j^t)B_j)-\rho_j(A_j)\rho_j(B)_j|
	\le C||A_j||_p||B_j||_pe^{-\gamma t}   $$
when $A_j,B_j\in{\cal C}^p(S_j)$.
\medskip
	We shall now see that if this holds for $j=0,\ldots,N$, then the flow $(f_\times^t)$ on $S=\times_0^NS_j$ defined by $f_\times^t({\bf x}_0,\ldots,{\bf x}_N)=(f_0^t{\bf x}_0,\ldots,f_N^t{\bf x}_N)$ is also exponentially mixing with respect to $\rho_\times=\times_0^N\rho_j$.  Indeed we shall prove that
$$	|\rho_\times((A\circ f_\times^t)B)-\rho_\times(A)\rho_\times(B)|
	\le(N+1)C||A||_p||B||_pe^{-\gamma t}\eqno{(A.1)}   $$
provided $A,B\in{\cal C}^p(S)$.
\medskip
	Define, for $j=0,\ldots,N+1$,
$$	\tilde A_j({\bf x}_j,{\bf x}_{j+1},\ldots,{\bf x}_N)=\int\rho_0(d{\bf x}_0)\cdots
	\rho_{j-1}(d{\bf x}_{j-1})A({\bf x}_0,\ldots,{\bf x}_j,\ldots,{\bf x}_N)   $$
and similarly for $\tilde B_j$.  By assumption, for all ${\bf x}'_{j+1},\ldots,{\bf x}'_N,{\bf x}''_{j+1},\ldots,{\bf x}''_N$, we have
$$	\big|\int\rho_j(d{\bf x}_j)\tilde A_j(f_j^t{\bf x}_j,{\bf x}'_{j+1},\ldots,{\bf x}'_N)
	\tilde B_j({\bf x}_j,{\bf x}''_{j+1},\ldots,{\bf x}''_N)   $$
$$	-(\tilde A_{j+1}({\bf x}'_{j+1},\ldots,{\bf x}'_N)
	\tilde B_{j+1}({\bf x}''_{j+1},\ldots,{\bf x}''_N)\big|
	\le C||A||_p||B||_pe^{-\gamma t}   $$
hence
$$	\big|\int\rho_j(d{\bf x}_j)\cdots\int\rho_N(d{\bf x}_N)
	\tilde A_j(f_j^t{\bf x}_j,\ldots,f_N^t{\bf x}_N)\tilde B_j({\bf x}_j,\ldots,{\bf x}_N)   $$
$$	-\int\rho_j(d{\bf x}_{j+1})\cdots\int\rho_N(d{\bf x}_N)
	\tilde A_{j+1}(f_{j+1}^t{\bf x}_{j+1},\ldots,f_N^t{\bf x}_N)
	\tilde B_{j+1}({\bf x}_{j+1},\ldots,{\bf x}_N)\big|   $$
$$	\le C||A||_p||B||_pe^{-\gamma t}   $$
Since 
$$	\int\rho_0(d{\bf x}_0)\cdots\int\rho_N(d{\bf x}_N)
	\tilde A_0(f_0^t{\bf x}_0,\ldots,f_N^t{\bf x}_N)\tilde B_0({\bf x}_0,\ldots,{\bf x}_N)
	=\rho_\times((A\circ f_\times^t)B)   $$
$$	\tilde A_{N+1}=\rho_\times(A)\qquad,\qquad\tilde B_{N+1}=\rho_\times(B)   $$
we obtain $(A.1)$.
\medskip
	The case of interest to us is when $(f_j^t)$ is an exponentially mixing Anosov flow with respect to the volume $\rho_j$ on $S_j$.  Then $(f_\times^t)$ is exponentially mixing with respect to the volume $\rho_\times$ on $S$, and therefore $f_\times^1$ is {\it rapidly mixing} in the sense of [8].  This means (roughly) that if $\ell_{u,\sigma}$ is a probability measure with $\eta$-H\"older density $\sigma$ on a local unstable manifold ${\cal W}^u$, and if $A$ has derivatives in the center direction which are $\eta$-H\"older in $S$, then $\ell_{u,\sigma}(A\circ f_\times^n)$ tends to $\rho_\times(A)$ faster than any $n^{-\kappa}$ (with $\kappa>0$) when $n\to\infty$.  To check that exponential mixing implies rapid mixing one can approximate $\delta$-measures on unstable disks (used to define $\ell_{u,\sigma}$) by smooth functions.  [I am indebted to Dmitry Dolgopyat for explaining this to me.]  In fact, spreading the mass of $\ell_{u,\sigma}$ by a small distance $\sim r$ along stable manifolds, then along center manifolds, we get a probability measure $\ell_\phi$ with $\epsilon$-H\"older density on $M$ such that $|\ell_{u,\sigma}(A\circ f_\times^n)-\ell_\phi(A\circ f_\times^n)|<C_1r^\eta$.  The exponent $\epsilon$ is determined from the H\"older exponent of $\sigma$ and of the stable foliations, and we choose $\epsilon<\eta$.  Smoothing $\phi\in{\cal C}^\epsilon$ to $\tilde\phi\in{\cal C}^p$ we have $||\phi-\tilde\phi||<C_2r^\epsilon$, so that $|\ell_\phi(A\circ f_\times^n)-\ell_{\tilde\phi}(A\circ f_\times^n)|<C'_2r^\epsilon$.  Smoothing $A$ to $\tilde A\in{\cal C}^p$ we have $||A-\tilde A||_0<C_3r^\eta$, so that $|\ell_{\tilde\phi}(A\circ f_\times^n)-\ell_{\tilde\phi}(\tilde A\circ f_\times^n)|<C'_3r^\eta$.  Assuming $\rho_\times(A)=0$ we may also assume $\rho_\times(\tilde A)=0$, and exponential mixing gives
$$	|\ell_{\tilde\phi}(\tilde A\circ f_\times^n)|=|\rho_\times(\tilde A\circ f_\times^n)\tilde\phi)|
	<C_4r^{-p}e^{-\gamma n}   $$
and thus
$$	|\ell_{u,\sigma}(A\circ f_\times^n)|
	<(C_1+C'_2+C'_3)r^\epsilon+C_4r^{-p}e^{-\gamma n}   $$
Taking $r$ such that
$$	r^{p+\epsilon}={p\over\epsilon}\cdot{C_4\over C_1+C'_2+C'_3} e^{-\gamma n}   $$
we have
$$	|\ell_{u,\sigma}(A\circ f_\times^n)|<C_5\exp(-\gamma{\epsilon\over p+\epsilon}n) $$ for large $n$.  In particular, $f_\times^1$ is rapidly mixing.
\medskip\noindent
{\bf Appendix B: proof of} (3.3).
\medskip
	The proof of theorem 1 in [8] involves the invariant splitting
$$   TS=E_\times^u\oplus E_\times^c\oplus E_\times^s   $$
associated with $(f_\times^t)$.  Here $E_\times^u=\oplus_0^NE_j^u$, $E_\times^c=\oplus_0^N E_j^c$, $E_\times^s=\oplus_0^NE_j^s$, where $E_j^u$ is the unstable vector bundle for $(f_j^t)$, $E_j^s$ the stable vector bundle, and $E_j^c$ the one-dimensional bundle in the direction of the flow, so that $(E_j^c)_{{\bf x}_j}$ is spanned by
$$   e_j({\bf x}_j)={d\over dt}f_j^t{\bf x}_j/||{d\over dt}f_j^t{\bf x}_j||  $$
The bundle $E_\times^s$ is H\"older continuous, but in general not smooth.  A smooth bundle $E_\times^{as}$, C$^0$-close to $E_\times^s$, is introduced.  The components $Z^u,Z^c,Z^{as}$ of a vector $Z$ will be taken with respect to the splitting $TS=E_\times^u\oplus E_\times^c\oplus E_\times^{as}$.  We shall use bundle maps $T^u,T^c,T^{as}: TS\mapsto TS$ such that $(T^u{\bf x}))Z=((T_{\bf x}f^1)Z)^u$, $(T^c({\bf x}))Z=((T_{\bf x}f^1)Z)^c$, $(T^{as}({\bf x}))Z=((T_{\bf x}f^1)Z)^{as}$.
\medskip
	Let now
$$	Y=({df^1\over d\lambda}\circ f^{-1})\big|_{\lambda=0}   $$
and define the vector field $V$ and the functions $a_j$ on $S$ by
$$	V({\bf x})=\sum_{n=0}^\infty(T^{as})^nY^{as}(f_\times^{-n}{\bf x})   $$
$$	Y^c+T^cV=\sum_{j=0}^Na_j({\bf x})e_j({\bf x})   $$
Then the functional $\omega$ such that $(3.2)$ holds  is given by Proposition 2.6 of [8]:
$$	\omega(A)=\rho_\times(\partial_VA)
	+\sum_{j=0}^N\sum_{n=0}^\infty\rho_\times((a_j\circ f_\times^{-n})\partial_{e_j}A)
	-\sum_{n=0}^\infty\rho_\times(([{\rm div}^u(Y^u+T^uV)]\circ f_\times^{-n})A)   $$
where ${\rm div}^u$ is the divergence with respect to the canonical density on ${\cal W}^u$.
\medskip
	We claim that we can transform the above formula to
$$	\omega(A)=-\sum_{k=0}^\infty\rho_\times([{\rm div}Y][A\circ f_\times^k])
	\eqno{(3.3)}   $$
To show that the right-hand sides are equal, we may replace $\rho_\times$  by its conditional measure $\rho_u$ on a local unstable manifold ${\cal W}^u$, i.e., integrate on ${\cal W}^u$ with respect to the canonical volume element.  The divergence ${\rm div}$ is with respect to the volume element $\rho_\times(d{\bf x})$, and along an unstable manifold it can be naturally factorized in volume elements along the $as,c$, and $u$ directions, so that we have
$$	-\rho_\times([{\rm div}Y][A\circ f_\times^k])
	=\rho_\times(Y^{as}\cdot\partial(A\circ f_\times^k))
	-\rho_\times([{\rm div}^{cu}Y^{cu}][A\circ f_\times^k])   $$
where
$$	\rho_\times(Y^{as}\cdot\partial(A\circ f_\times^k))
	=\rho_\times((Tf_\times^1)Y^{as}\cdot\partial(A\circ f_\times^{k-1}))   $$
$$	=\rho_\times((T^{as}Y^{as})\cdot\partial(A\circ f_\times^{k-1}))
	-\rho_\times([{\rm div}^{cu}(T^{cu}Y^{as})][A\circ f_\times^{k-1}])=\cdots   $$
$$	=\rho_\times(((T^{as})^kY^{as})\cdot\partial A)
	-\sum_{\ell=1}^k\rho_\times([{\rm div}^{cu}(T^{cu}(T^{as})^{\ell-1}Y^{as}]
	[A\circ f_\times^{k-1}])   $$
Using (6) and Lemma B.1 of [8] we see that the sum
$$	\sum_{k,\ell}\rho_\times([{\rm div}^{cu}(T^{cu}(T^{as})^{\ell-1}Y^{as}]
	[A\circ f_\times^{k-1}])   $$
converges absolutely, so that
$$	-\sum_{k=0}^\infty\rho_\times([{\rm div}Y][A\circ f_\times^k])
	=\rho_\times(V\cdot\partial A)
	-\sum_{k=0}^\infty\rho_\times([{\rm div}^{cu}(Y^{cu}+T^{cu}V][A\circ f_\times^k]) $$
and the right-hand side is equal to the expression for $\omega(A)$ of [8] reproduced above.  We have thus proved $(3.3)$.
\medskip\noindent
{\bf Appendix C: proof of} (3.1).
\medskip
	In our case
$$	{\rm div}\,Y({\bf x})=\lambda\,{\rm div}\int_0^1dt\,(T_{f^{-t}{\bf x}}f^t)\tilde X
	=\lambda\,\int_0^1dt\,({\rm div}\tilde X)(f^{-t}{\bf x})   $$
where $\tilde X\in TS$ has components
$$	\pmatrix{X_{ju}-\alpha_jp_{ju}\cr0\cr}   $$
Therefore
$$	\omega(A)=-\sum_{k=0}^\infty\rho_\times([{\rm div}Y][A\circ f_\times^k])
	=-\lambda\sum_{k=0}^\infty\int_0^1dt\,
	\rho_\times([{\rm div}\tilde X][A\circ f_\times^{k+t}])   $$
$$	=-\lambda\int_0^\infty dt\,\rho_\times([{\rm div}\tilde X][A\circ f_\times^t])   $$
One proves readily the formula $\partial_{{\bf p}_j}\cdot(\alpha_jp_j)=(n-1)\alpha_j$, from which we obtain
$$	-{\rm div}\tilde X=(n-1)\sum_{j=0}^N\alpha_j   $$
hence
$$	\omega(A)
=\lambda(n-1)\int_0^\infty dt\,\rho_\times((\sum_{j=0}^N\alpha_j)(A\circ f_\times^t))   $$
i.e., we have proved $(3.1)$.\qed
\medskip\noindent
{\bf References.}
\medskip
[1] W. Aschbacher, V. Jak\v si\'c, Y. Pautrat, and C.-A. Pillet.  ``Introduction to nonequilibrium quantum statistical mechanics'', pp. 1-66 in {\it Open Quantum Systems III Recent developments.}  Lecture Notes in Mathematics {\bf 1882}, Springer, Berlin, 2006.

[2] V. Baladi and D. Smania.  ``Linear response for smooth deformations of generic nonuniformly hyperbolic unimodal maps.''  To be published.

[3] C. Bonatti, L. Diaz, and M. Viana.  {\it Dynamics beyond uniform hyperbolicity.} Sprin\-ger, Berlin, 2005.

[4] F. Bonetto, J. Lebowitz, and L. Rey-Bellet.  ``Fourier's law: a challenge for theorists.'' pp. 128-150 in {\it Mathematical Physics 2000}, A. Fokas, A. Grigoryan, T. Kibble and B. Zegarlinsky (eds), Imperial College, London, 2000. 

[4a] F. Bonetto, J.L. Lebowitz, and J. Lukkarinen.  ``Fourier's law for a harmonic crystal with self-consistent stochastic reservoirs.''  J. Statist. Phys. {\bf 116},783-813(2004).

[4b] F. Bonetto, J.L. Lebowitz, J. Lukkarinen, and S. Olla.  ``The heat conduction and entropy production in anharmonic crystals with self-consistent stochastic reservoirs.''  J. Statist. Phys. {\bf 134},1097-1119(2009).

[5] O. Butterley and C. Liverani  ``Smooth Anosov flows: correlation spectra and stability."  J. Modern Dynamics {\bf 1},301-322(2007).

[6] D. Dolgopyat.  ``Decay of correlations in Anosov flows.''  Ann. of Math. {\bf 147},357-390(1998).

[7] D. Dolgopyat.  ``Prevalence of rapid mixing in hyperbolic flows.''  Ergod. Th. and Dynam. Syst. {\bf 18},1097-1114(1998).  ``Prevalence of rapid mixing-II: topological prevalence''  Ergod. Th. and Dynam. Syst. {\bf 20},1045-1059(2000).

[8] D. Dolgopyat.  ``On differentiability of SRB states for partially hyperbolic systems"  Invent. Math. {\bf 155},389-449(2004).

[8a] D. Dolgopyat and C. Liverani.  ``Energy transfer in a fast-slow Hamiltonian system.''  Preprint.

[9] J.-P. Eckmann, C.-A. Pillet, and L. Rey-Bellet.  ``Non-equilibrium statistical mechanics of anharmonic chains coupled to two baths at different temperatures.''  Commun. Math. Phys. {\bf 201},657-697(1999).

[10] J.-P. Eckmann and L.-S. Young.  ``Temperature profiles in Hamiltonian heat conduction.''  Europhys. Lett. {\bf 68},790-796(2004).

[11] J.-P. Eckmann and L.-S. Young.  ``Nonequilibrium energy profiles in Hamiltonian for a class of 1-D models.''  Commun. Math. Phys. {\bf 262},237-267(2006).

[12] D.J. Evans and G.P. Morriss.  {\it Statistical mechanics of nonequilibrium fluids.}  Academic Press, New York, 1990.

[13] G. Gallavotti and E.G.D. Cohen.  ``Dynamical ensembles in nonequilibrium statistical mechanics.''  Phys. Rev. Letters {\bf 74},2694-2697(1995); ``Dynamical ensembles in stationary states.'' J. Statist. Phys. {\bf 80},931-970(1995).

[14] M. Hirsch, C.C. Pugh, and M. Shub.  {\it Invariant manifolds.}  Lect. Notes in Math. {\bf 583} Springer, Berlin, 1977.

[15] W.G. Hoover.  {\it Molecular dynamics.}  Lecture Notes in Physics {\bf 258}.  Springer, Heidelberg, 1986.

[16] A. Katok, G. Knieper, M. Pollicott, and H. Weiss.  "Differentiability and analyticity of topological entropy for Anosov and geodesic flows."  Invent. Math. {\bf 98},581-597(1989).

[17] Ya. B. Pesin and Ya. G. Sinai.  ``Gibbs measures for partially hyperbolic attractors.''  Ergod. Th. and Dynam. Syst. {\bf 2},417-438(1982).

[18] D. Ruelle.  ``Differentiation of SRB states.''  Commun. Math. Phys. {\bf 187},227-241(1997); ``Correction and complements.''  Commun. Math. Phys. {\bf 234},185-190(2003).

[19] D. Ruelle.  ``Smooth dynamics and new theoretical ideas in nonequilibrium statistical mechanics.''  J. Statist. Phys. {\bf 95},393-468(1999).

[20] D. Ruelle.  ``Differentiation of SRB states for hyperbolic flows."  Ergod. Theor. Dynam. Syst. {\bf 28},613-631(2008).

[21] D. Ruelle.  ``Singularities of the susceptibility of an SRB measure in the presence of stable-unstable tangencies.''  Phil. Trans. R. Soc. A{\bf 369},482-493(2011).

[22] N. G. van Kampen.  ``The case against linear response theory.''  Phys. Norv. {\bf 5},279-284(1971).

[23] L.-S. Young.  ``What are SRB measures, and which dynamical systems have them?''  J. Statist. Phys. {\bf 108},733-754(2002).

\end